# Out-of-plane structural flexibility of phosphorene

Gaoxue Wang[1*], G. C. Loh[2], Ravindra Pandey[1*], and Shashi P. Karna[3]

[1]Department of Physics, Michigan Technological University, Houghton, Michigan 49931, USA

[2]Institute of High Performance Computing, 1 Fusionopolis Way, #16-16 Connexis, Singapore 138632

[3]US Army Research Laboratory, Weapons and Materials Research Directorate, ATTN: RDRL-WM, Aberdeen Proving Ground, MD 21005-5069, USA

(August 31, 2015)

Email:  gaoxuew@mtu.edu
        pandey@mtu.edu

## Abstract

Phosphorene has been rediscovered recently, establishing itself as one of the most promising two dimensional group-V elemental monolayers with direct band gap, high carrier mobility, and anisotropic electronic properties. In this paper, the buckling and its effect on the electronic properties in phosphorene are investigated by using molecular dynamics simulations and complemented by density functional theory calculations. We find that phosphorene shows superior out-of-plane structural flexibility along the armchair direction, which allows the formation of buckling with large curvatures, while the buckling along the zigzag direction will break its structure integrity at large curvatures. The semiconducting and direct band gap nature are retained with buckling along the armchair direction; the band gap decreases and transforms to an indirect band gap with buckling along the zigzag direction. The structural flexibility and electronic robustness along the armchair direction facilitate the fabrication of devices with complex shapes, such as folded phosphorene and phosphorene nano-scrolls, thereby offering new possibilities for the application of phosphorene in flexible electronics and optoelectronics.

**Keywords:** phosphorene, buckling, structural flexibility, molecular dynamics, density functional theory

**1.0 Introduction**

Buckling is one of the most important mechanical phenomena in two dimensional (2D) materials including graphene which has elicited broad scientific interests[1-4]. Graphene possesses a high in-plane Young's modulus with *sp*$^2$ bonded carbon atoms[5], while it can easily be warped in the out-of-plane direction enabling folding[6], bending[7], corrugating[8] or wrinkling[9] without loss of its structural integrity[10]. This structural flexibility facilitates the fabrication of graphene-based complex structures with distinct functionalities[9, 11-17]. Furthermore, buckling often appears in graphene grown from chemical vapor deposition (CVD)[15, 18, 19], and can be controlled via thermally activated shape-memory polymer substrates[16]. Similar buckling has also been observed in other 2D materials, such as hexagonal boron nitride (*h*-BN)[20], and molybdenum disulphide (MoS$_2$)[21].

The monolayer form of black phosphorus, also known as phosphorene, has drawn considerable attention recently as a novel 2D semiconducting material[22, 23]. High-quality phosphorene has been exfoliated by the mechanical[23, 24] or liquid method[25] with a fundamental direct band gap[26]. Moreover, the carrier mobility in few layers of phosphorene could reach 1000 cm$^2$/(V·s)[22], which is higher than that of 200 cm$^2$/(V·s) in MoS$_2$[27]. Other fascinating properties, such as the anisotropic conductance[28], fast optical response[29], and superior mechanical property[30] make phosphorene a promising candidate for electronics devices based on 2D materials. The mechanical properties of phosphorene under tensile strains have been investigated using both density functional theory (DFT) calculations[30] and classical molecular dynamics (MD) simulations[31]. The formation of ripples in phosphorene under a compressive strain was also investigated using DFT calculations[32]. However, the previous DFT study on the ripples[32] was unable to capture the dynamical essence of the phosphorene membrane at finite temperatures, and the ripples were limited to small surface curvatures.

In this paper, the buckling and its effect on the electronic properties of phosphorene are studied by using classical MD simulations complemented by calculations based on DFT. The MD simulations allow us to investigate the dynamical process of buckling at large scale with modest computational resources. For the buckled configurations obtained by MD simulations, the electronic properties are further determined by DFT calculations. The calculated results find that the buckling behavior of phosphorene can be described by Euler's buckling rule. More importantly, phosphorene shows superior out-of-plane structural flexibility along the armchair direction. The semiconducting and direct band gap nature are retained with buckling at large curvatures, which facilitates its application in flexible electronics and optoelectronics.

**2.0 Computational method**

The classical MD simulations were performed using the large-scale atomic/molecular massively parallel simulator (LAMMPS)[33] code. In phosphorene, the interatomic interactions were characterized by the Stillinger-Weber (SW) potential[34]. The SW potential has been previously parameterized to correctly describe the mechanical properties of phosphorene[31]. In MD simulations, phosphorene membranes with different dimensions were considered and the periodic boundary conditions were applied to both the armchair and the zigzag directions.

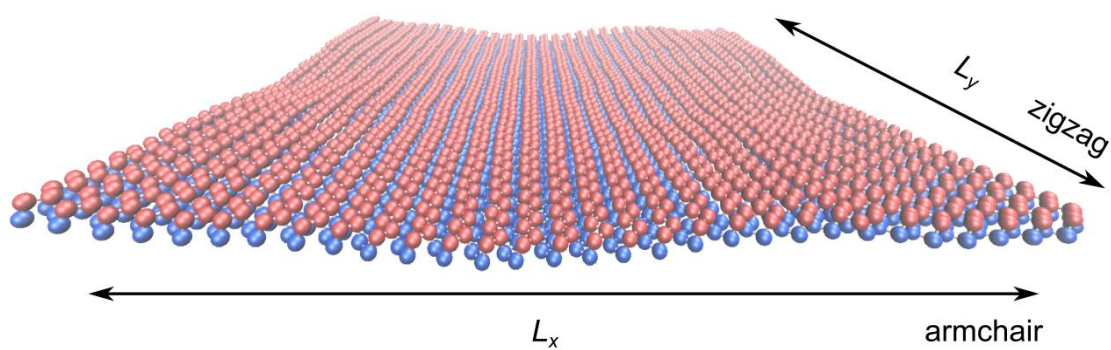

*Figure 1. Snapshots of phosphroene at a thermally stable state at 300 K. $L_x$ is the supercell size along the armchair direction, and $L_y$ is the size along the zigzag direction.*

Figure 1 shows one snapshot of phosphorene membrane at a thermally stable state. Initially, the structure of phosphorene membrane was minimized using the SW potential. After minimization, the monolayer was equilibrated to a thermally stable state with the NVT (constant particle number, constant volume, and constant temperature) ensemble for 250 *ps*, followed by the NPT (constant particle number, constant pressure, and constant temperature) ensemble for 250 *ps*. After equilibration, phosphorene was compressed in either the armchair or zigzag direction with a strain rate of $10^{-4}$ *ps*$^{-1}$, while the stress in the lateral direction was allowed to relax. To eliminate the layer-layer interaction, simulation boxes with thickness of 10 nm were used. The temperature was set to 0.1 K or 300 K, the pressure to 0 bar, and the time step was set to 0.5 *fs*. The VMD[35] software package was used to visualize the simulation results. The strain is defined as the change of supercell size along the armchair or the zigzag direction ($\varepsilon = \frac{\Delta L_x}{L_x}, or \frac{\Delta L_y}{L_y}$).

Due to the structural anisotropy of phosphorene as shown in Figure 1, the buckling along the armchair and the zigzag direction is expected to be different. Thus, different samples with variable sizes as listed in Table S1 (see supplementary information) were used to simulate the buckling. The size of supercell along strain direction was varied from ~60 to 160 Å, while the size in the lateral direction was close to ~130 Å.

The electronic properties of the buckled phosphorene were obtained by DFT calculations using the norm-conserving Troullier-Martins pseudopotential as implemented in SIESTA[36]. The Perdew-Burke-Ernzerhof (PBE)[37] exchange correlation functional and a double-$\zeta$ basis including polarization orbitals were employed. Supercells of (30×1) and (1×30) were used for buckling along the armchair and the zigzag direction, respectively. The reciprocal space was sampled by a grid of (5×1×1) or (1×5×1) *k* points in the Brillouin zone, respectively. The buckled configurations with different curvatures obtained from the snapshots of LAMMPS simulations at 0.1 K were taken as the initial configurations for DFT calculations. The energy convergence was set to $10^{-5}$ eV for electronic self-consistency steps. The mesh cutoff energy was 500 Ry (i.e. 6802 eV). The geometry optimization was considered to converge when the residual force on each atom was smaller than 0.01 eV/Å. The atoms were allowed to relax during the structural optimization, while the size of the supercell was fixed. Note that lattice constants obtained by the SW potential along the armchair and the zigzag direction (4.38 Å and 3.31 Å, respectively) are in agreement to those obtained from the DFT calculations (4.57 Å and 3.31 Å, respectively).

**3.0 Results and discussion**

**3.1 Buckling of phosphorene under a compressive strain**

Figure 2 shows the structural evolution of phosphorene with the applied compressive strain ($\varepsilon$) along the armchair and the zigzag directions at 300 K. With small $\varepsilon$, phosphorene maintains a flat surface with small ripples due to thermal vibrations. Buckling structure forms with slightly larger strains applied along both directions. Increasing the magnitude of $\varepsilon$ deforms phosphorene with the enhancement of the buckling height in the out-of-plane direction. Interestingly, the structural integrity of phosphorene is preserved under a large strain along the armchair direction, while the bonds are broken at a large strain along the zigzag direction (Figure 2(b)).

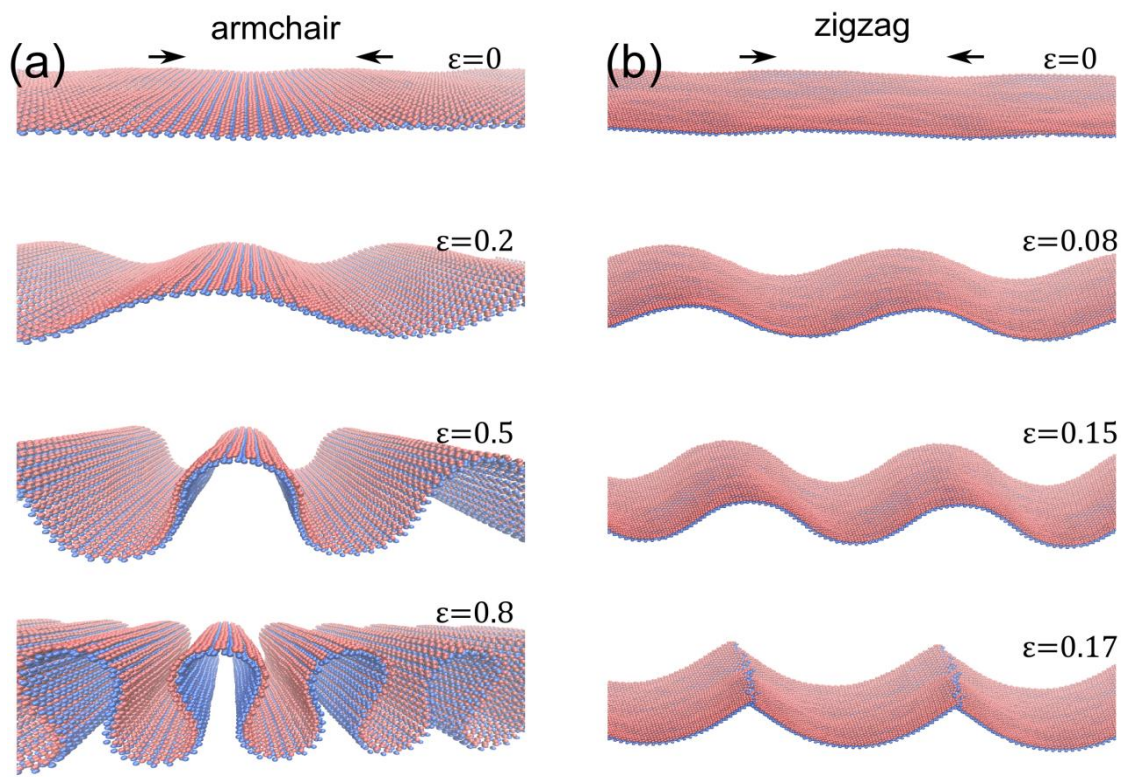

*Figure 2. Snapshots of phosphroene (cell size=(30×40)) under in-plane compressive strain ($\varepsilon$) at 300 K: (a) strain along armchair direction, (b) strain along zigzag direction. The structures are shown in periodic manner along strain direction.*

The difference in buckling along the armchair and the zigzag direction stems from its structural anisotropy. As seen in Figure 1, the phosphorous atoms are arranged in a puckered lattice along the armchair direction. The puckered structure could accommodate external strains by changing the pucker angle without much distortion of the bond length, thereby giving rise to its structural flexibility. This is also the origin of the superior mechanical properties of phosphorene under tensile strains[30]. However, in the zigzag direction, the phosphorus atoms are bonded into a zigzag chain like structure (Figure 1) which is less flexible.

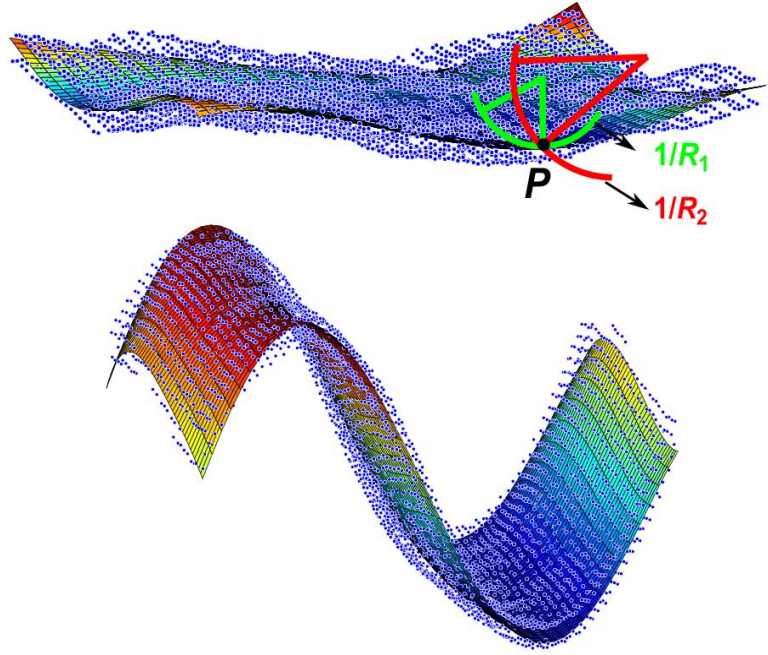

*Figure 3. Polynomial fitting of phosphorene surface. The blue dots are phosphorus atoms. The mean curvature at each point P is calculated on the fitted surface. $1/R_1$ and $1/R_2$ are the principle curvatures at P point. The mean curvature is defined as $\frac{1}{2}\left(\frac{1}{R_1}+\frac{1}{R_2}\right)$ at each point.*

To quantitatively describe the buckling behavior, we calculate the curvature of phosphorene membrane as illustrated in Figure 3. Since phosphorene has two sub-layers of phosphorus atoms, a polynomial fitting of the surface yields the principle curvatures at each point of the surface. The mean curvature at each point (P) on the surface is defined as half of the sum of the principle curvatures, $\frac{1}{2}(\frac{1}{R_1}+\frac{1}{R_2})$, where $\frac{1}{R_1}$ and $\frac{1}{R_2}$ are the principle curvatures.

Figure 4 shows the change of maximum mean curvature of phosphorene under a compressive strain along the armchair and the zigzag directions. It has distinct trends for the cases of $\varepsilon < \varepsilon_c$ and $\varepsilon > \varepsilon_c$, where $\varepsilon_c$ is the critical strain for the formation of buckling as illustrated by the vertical dashed line in the inset.

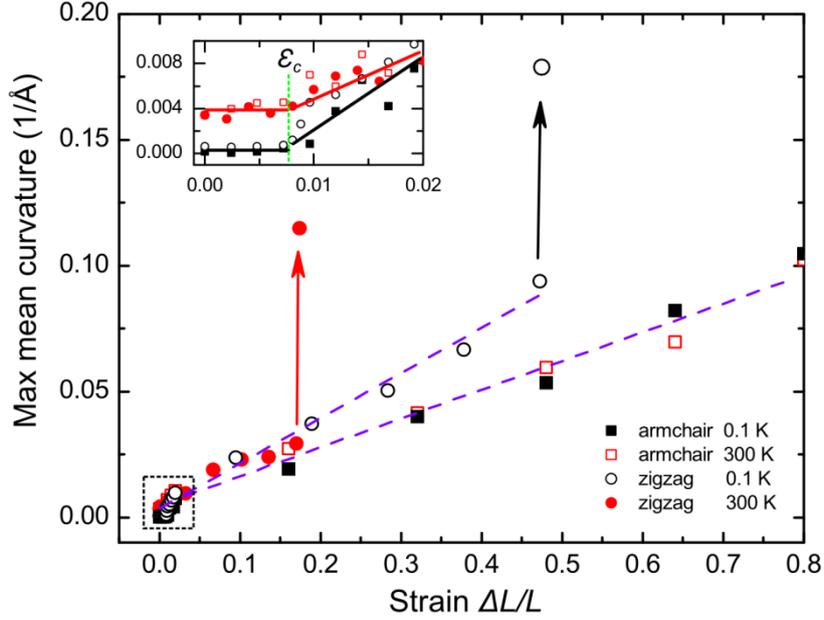

*Figure 4. Maximum mean curvature of phosphorene (cell size=(30×40)) under compressive strains along the armchair (square) and the zigzag directions (circle) at a temperature of 0.1 K (black) or 300 K (red). The solid lines are guides to the eye. The arrow represents the break of the structure along zigzag direction with an abrupt increase of the maximum mean curvature. The inset is the zoomed in plot in the small strain region in the dashed box, the dashed line in the inset corresponds to the buckling critical strain.*

For $\varepsilon < \varepsilon_c$, as shown in the inset of Figure 4, the maximum mean curvature is almost unchanged along both the armchair and the zigzag directions, which corresponds to the elastic response of the membrane to external strain. During this process, the surface keeps almost flat with small vibrations due to thermally excited ripples. For $\varepsilon > \varepsilon_c$, the maximum mean curvature starts to increase, which corresponds to the formation of buckling. The mean curvature increases linearly with the strain on phosphorene. The buckling critical strain $\varepsilon_c$ is ~0.007 along armchair and zigzag directions for the sample with supercell size of (30×40).

The buckling curvature along the armchair direction linearly increases with $\varepsilon$ up to 0.8 inducing the formation of folded phosphorene without breaking the structural integrity (see also in Figure 2(a)), which suggests its flexibility along the armchair direction. In the zigzag direction, an abrupt increase appears in the maximum mean curvature curve (illustrated by arrows in Figure 4), which corresponds to the breaking of the structure with abrupt release of stress (see also in Figure 2(b)). The breaking strain of the structure at 0.1 K is 0.47, which decreases to 0.17 at 300 K. Therefore, a large strain along the zigzag direction will break the structural integrity of phosphorene.

According to Euler's buckling theory[38], a thin plate will experience buckling due to a compressive strain applied on it. The buckling critical strain is an inverse quadratic function of the length of the plate, $\varepsilon_c \propto -\frac{1}{L^2}$, where $L$ is the length of the plate. The length dependence of buckling critical strain for various samples is summarized in Figure 5. The critical strain decreases with the increase of the sample size in both the armchair and the zigzag directions, which can be well fitted with Euler's buckling rule, $\varepsilon_c \propto -\frac{1}{L^2}$.

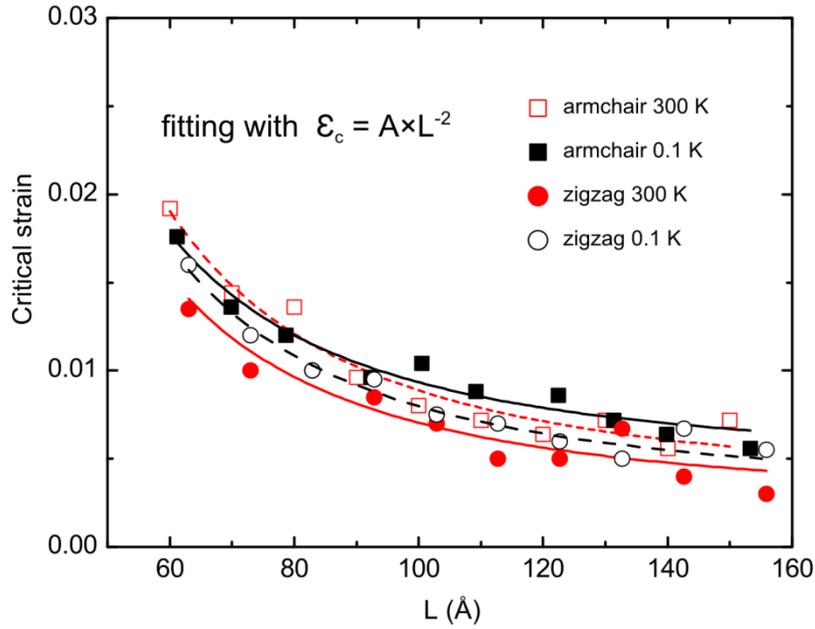

*Figure 5. Buckling cricital strain vs the size of the simulation sample. The lines are fitted curve according to the Euler's buckling theory.*

**3.2 Electronic properties of buckled phosphorene**

To investigate the electronic properties of buckled phosphorene, DFT calculations were performed on the buckled structures with various curvatures obtained at the classical MD simulations. Note that the buckled structures at low temperature were chosen to enable the fast convergence during DFT calculations. Strain free phosphorene has a direct band gap of ~1 eV in our calculations, which agrees with previous theoretical values[22, 23, 39-41].

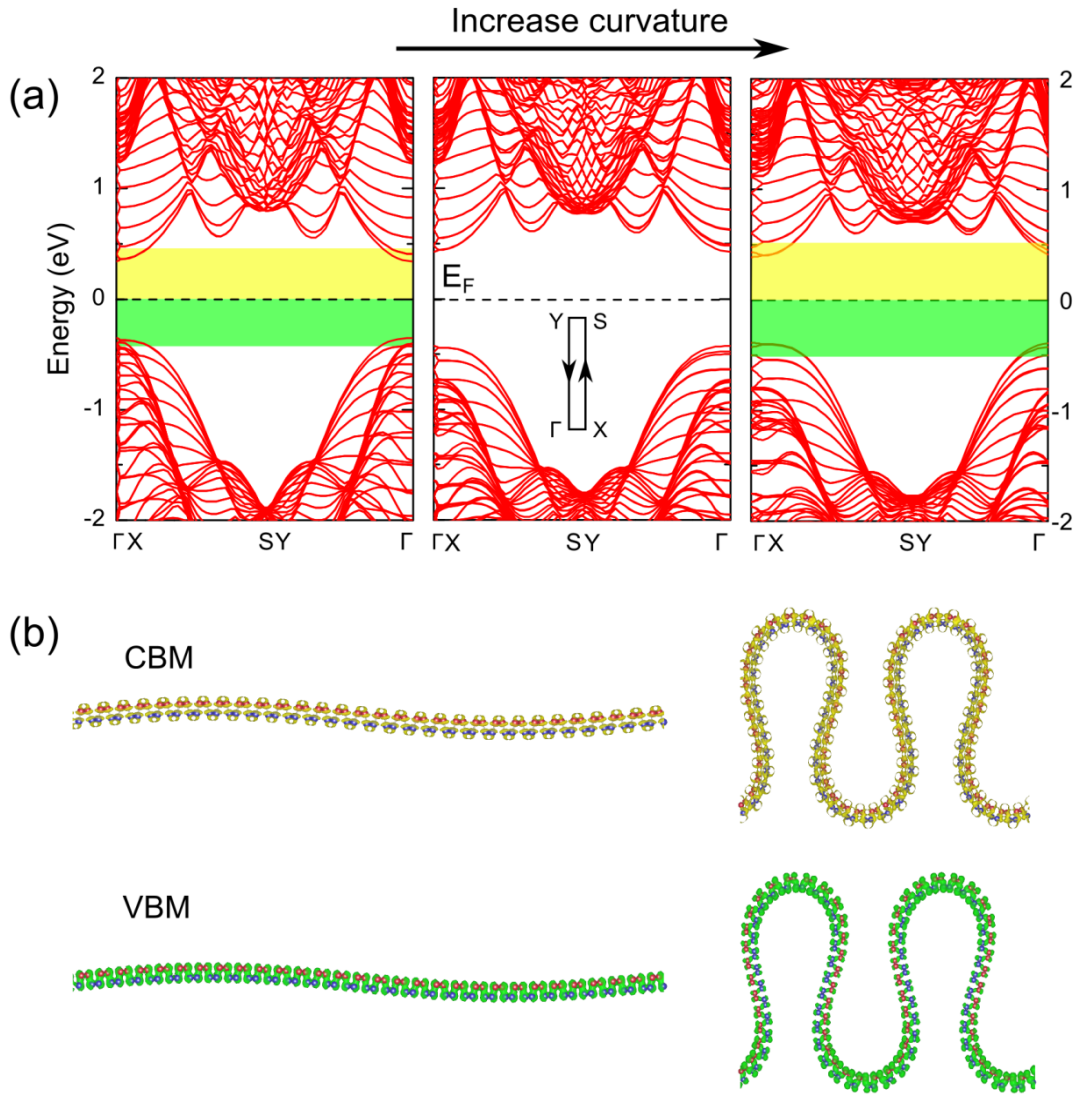

*Figure 6. Electronic properties of phosphorene with buckling along armchair direction: (a) band structures at different curvature, (b) charge density at VBM and CBM. The inset is the Brillouin zone.*

Figure 6 shows the band structures and charge density at conduction band minimum (CBM) and at valence band maximum (VBM) with buckling along the armchair direction. Low buckled phosphorene has a direct band gap at Γ. The charge density at VBM and CBM are evenly distributed over the surface as seen in Figure 6 (b). The semiconducting property, direct band gap, and evenly distributed charge density are retained in largely buckled phosphorene suggesting the electronic robustness of phosphorene to the buckling along the armchair direction.

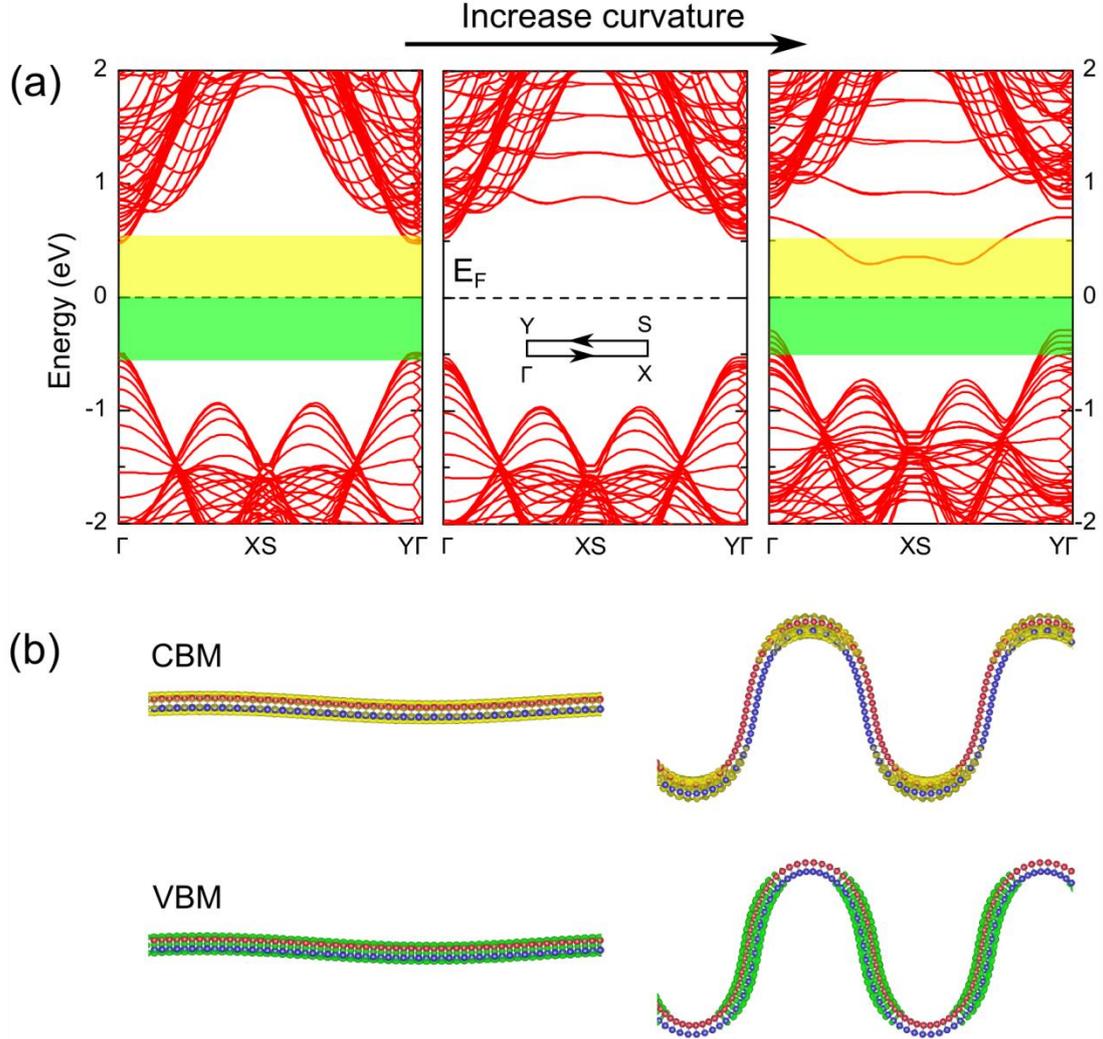

*Figure 7. Electronic properties of phosphorene with buckling along zigzag direction: (a) band structures at different curvatures, (b) charge density at VBM and CBM. The inset is the Brillouin zone.*

As seen in Figure 7, low buckled structure along the zigzag direction has a direct band gap, the charge density at VBM and CBM is evenly distributed over the surface as expected. By increasing the curvature of the buckling, some conduction states approach Fermi level decreasing the band gap and thereby inducing a direct-indirect band gap transition. We observe unevenly distributed charge density at VBM and CBM at large curvature. The conduction states contributing to the decrease of the band gap comes from the convex region of the buckled surface due to the accumulated local strains in these regions (Figure 7 (b)). Therefore, compared to the buckling in the armchair direction, buckled phosphorene along the zigzag direction is less robust in terms of the structural and electronic properties of a candidate two-dimensional material for device applications.

**Summary**

In summary, we investigate the buckling in phosphorene under compressive strains by using classical MD simulation combined with first-principles calculations. A few interesting results are obtained from present study. (i) Buckling will form in phosphorene under a compressive strain along the armchair and the zigzag direction. The buckling critical strain satisfies the Euler's buckling theory. (ii) Phosphorene shows superior out-of-plane structural flexibility along the armchair direction, which allows the formation of buckling with large curvature; the buckling along the zigzag direction may break the structural integrity at large curvatures. (iii) The semiconducting and direct band gap nature of phosphorene are robust with the formation of buckling along the armchair direction; while buckling with large curvature along the zigzag direction will induce a direct to indirect band gap transition. The out-of-plane structural flexibility and electronic robustness of phosphorene along the armchair direction allow the fabrication of phosphorene based devices with complex shapes, such as folded structures and nano-scrolls in Figure S1 (see supplementary information). Our results contribute to the understanding of mechanical properties of phosphorene, and guide the design of phosphorene-based devices for flexible electronics and optoelectronics.

**Acknoledgements**

RAMA and Superior, high performance computing clusters at Michigan Technological University, were used in obtaining results presented in this paper. Supports from Dr. S. Gowtham are acknowledged. Financial support from ARL W911NF-14-2-0088 is acknowledged.

Table of Contents (TOC) Graphics

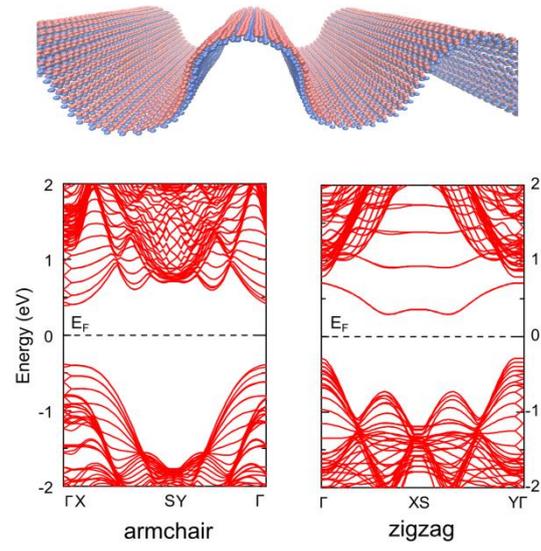

*Out-of-plane structural flexibility of phosphorene*

Gaoxue Wang[1*], G. C. Loh [2], Ravindra Pandey[1*], and Shashi P. Karna[3]

[1]Department of Physics, Michigan Technological University, Houghton, Michigan 49931, USA

[2]Institute of High Performance Computing, 1 Fusionopolis Way, #16-16 Connexis, Singapore 138632

[3]US Army Research Laboratory, Weapons and Materials Research Directorate, ATTN: RDRL-WM, Aberdeen Proving Ground, MD 21005-5069, USA

(August 31, 2015)

Email:  gaoxuew@mtu.edu
         pandey@mtu.edu

Table S1. The size of the supercell in terms of $L_x$ and $L_y$ used for MD calculations. The unit is Å.

|  | supercell | 14×40 | 16×40 | 18×40 | 21×40 | 23×40 | 25×40 | 28×40 | 30×40 | 32×40 | 35×40 |
|---|---|---|---|---|---|---|---|---|---|---|---|
| Armchair | $L_x$ | 61.2 | 69.9 | 79.0 | 92.3 | 100.5 | 109.1 | 122.4 | 131.3 | 139.8 | 153.3 |
|  | $L_y$ | 132.6 | 132.6 | 132.6 | 132.6 | 132.7 | 132.6 | 132.6 | 132.6 | 132.6 | 132.6 |
|  | supercell | 30×19 | 30×22 | 30×25 | 30×28 | 30×31 | 30×34 | 30×37 | 30×40 | 30×43 | 30×47 |
| Zigzag | $L_x$ | 131.2 | 131.2 | 131.4 | 131.8 | 131.1 | 131.1 | 131.2 | 131.3 | 131.2 | 131.3 |
|  | $L_y$ | 63.0 | 72.9 | 82.9 | 92.9 | 102.8 | 112.7 | 122.7 | 132.6 | 142.6 | 155.9 |

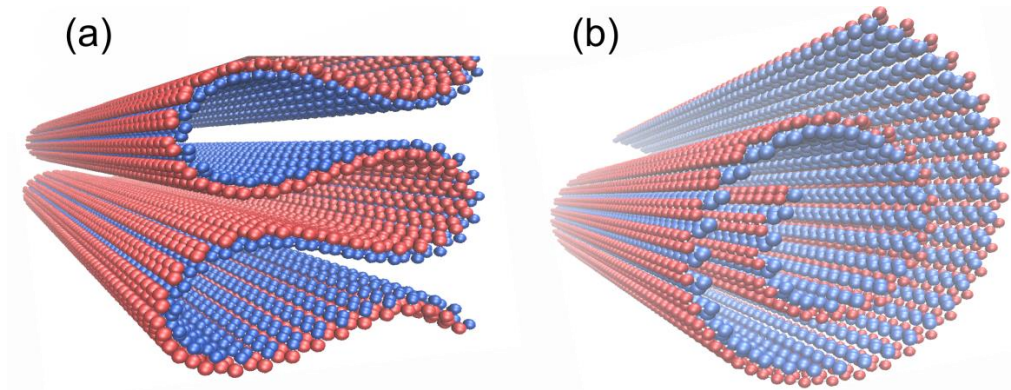

Figure S1. Folded phosphorene (a), and phosphorene nano-scroll (b).